\documentclass[aps,prd,twocolumn,showpacs,endfloats]{revtex4}
\usepackage{epsfig}
\usepackage{graphics}
\begin{document}
\preprint{\vbox{}}
\def\vp{{\bf p}}
\def\ko{K^0}
\def\kb{\bar{K^0}}
\def\al{\alpha}
\def\ab{\bar{\alpha}}
\def\besub{\begin{subequations}}
\def\eesub{\end{subequations}}
\def\laball{\label{allequations}}
\def\be{\begin{equation}}
\def\ee{\end{equation}}
\def\bea{\begin{eqnarray}}
\def\eea{\end{eqnarray}}
\def\non{\nonumber}
\def\lab{\label}
\def\la{\langle}
\def\ra{\rangle}
\def\epp{\epsilon^{\prime}}
\def\vep{\varepsilon}
\def\to{\rightarrow}
\def\up{\uparrow}
\def\dw{\downarrow}
\def\ms{\overline{\rm MS}}
\def\ums{{\mu}_{_{\overline{\rm MS}}}}
\def\u{\mu_{\rm fact}}

\def\pr{{Phys. Rev.}~}
\def\ijmp{{Int. J. Mod. Phys.}~}
\def\jp{{J. Phys.}~}
\def\mpl{{Mod. Phys. Lett.}~}
\def\prp{{Phys. Rep.}~}
\def\prl{{Phys. Rev. Lett.}~}
\def\pl{{Phys. Lett.}~}
\def\np{{Nucl. Phys.}~}
\def\ppnp{{Prog. Part. Nucl. Phys.}~}
\def\zp{{Z. Phys.}~}
\def\epj{{Eur. Phys. J.}~}

\title{Charm Antiquark and Charm Quark in the Nucleon$^*$\\}

\author{Xiaotong Song}
\email{xs3e@virginia.edu}
\par
\thanks{\\
$^*$ Report No. INPP-UVA-2001-04 (September 18, 2001)
} 

\affiliation{Institute of Nuclear and Particle Physics\\
Jesse W. Beams Laboratory of Physics\\
Department of Physics, University of Virginia\\
Charlottesville, VA 22904, USA\\}

\begin{abstract}
We estimate the intrinsic charm contributions to the quark flavor and 
spin observables of the nucleon in the SU(4) quark meson fluctuation 
model. In this model, the charm or anticharm reside in the charmed 
mesons created by the nonperturbative quantum quark-meson fluctuations. 
The intrinsic charm content in the proton, $2\bar c/\sum(q+\bar q)
\simeq 0.011\pm 0.008$, is almost one order of magnitude smaller than 
the intrinsic strange content. The intrinsic charm helicity is also 
small and negative, $\Delta c\simeq -(0.009\pm 0.006)$. The fraction 
of the total quark helicity carried by the charm is $|\Delta c/\Delta
\Sigma|\simeq 0.021\pm 0.014$. The ratio of the charm with positive 
helicity to that with negative helicity is $c_\up/c_\dw=35/67$. For 
the intrinsic strange component, one has $s_\up/s_\dw\simeq 7/13$. 
A detail comparison of our predictions with data and other models or 
analyses is given. The intrinsic charm contribution to the Ellis-Jaffe 
sum rule is also discussed.
\end{abstract}
\par
\par
\par

\par
\par

\maketitle

\leftline{\bf 1. Introduction}
\par

As suggested by many authors long time ago \cite{stan81,dg77}, there are 
so called {\it `intrinsic'} heavy quark components in the proton wave 
function. The intrinsic heavy quarks are created from the quantum 
fluctuations associated with the bound state hadron dynamics. They
exist in the hadron over a long time independent of any external 
probe momentum transfered in the collision. The probability of finding 
the intrinsic heavy quarks in the hadron is completely determined by 
nonperturbative mechanisms. On the other hand, the {\it extrinsic} 
heavy quarks are created on a short time scale in association with a 
large transverse momentum reaction and their distributions can be 
derived from QCD bremsstrahlung and pair production processes, which 
lead to standard QCD evolution. The study of nonperturbative mechanism 
$gg\to q\bar q$ \cite{stan81} shows that the intrinsic heavy quark 
contribution scales as $1/m_Q^2$, where $m_Q$ is the mass of the 
heavy quark, the probability of finding the intrinsic bottom is 
expected to be much smaller, hence we will only consider the intrinsic 
charm (IC) component, e.g. $|uudc\bar c>$, in the nucleon.

Many works on intrinsic charm are based on models of the nucleon (see 
e.g. MIT bag model \cite{dg77}, NJL model \cite{hk94}, meson 
cloud model \cite{stan81,nnnt96}, instanton model \cite{blos98,amt98} 
and others \cite{pst99}, etc.) or combination of the IC hypothesis 
and analysing the existing DIS data on charm production to obtain some 
information of the intrinsic charm content (see e.g. 
\cite{hm83,hsv96,vb96,it97,gol00,bs91,hz97,gv97,mt97,smt99,bn99}).
As pointed out in \cite{gv97} that a non-perturbative IC component
was unable to explain enhancement in the $e^+p$ neutral current cross 
section at HERA kinematics, but HERA data may provide a good test of 
the intrinsic charm if the charm component of $F_2$ is directly 
isolated and no new physics exists. Another discussion 
given in \cite{mt97} suggested a nonsymmetric charm distribution, for 
which $\bar c$ is much harder than $c$, which may account for the HERA
anomaly, a significant enhancement of cross sections at HERA at large
$x$ and $Q^2$ \cite{h196,zeus97}. However, the most recent analysis 
given in \cite{smt99}, which includes both the perturbative photon-gluon 
fusion process and nonperturbative intrinsic charm contribution, found 
no conclusive evidence for the intrinsic charm and only provided an upper 
bound of around 0.4$\%$. Therefore what is the effect of charm in the
structure of the proton is still an open and interesting question. A
comprehensive review on the nucleon sea, which includes light quark, 
intrinsic strange and charm, can be found in \cite{vogt00}.

The chiral quark model or more precisely the quark meson fluctuation 
model (in SU(2) version) was first suggested in \cite{ehq92} to explain 
the sea flavor asymmetry, $\bar d-\bar u>0$. The model was extended to 
SU(3) without and with SU(3) symmetry breaking to describe the 
quark spin structure (see e.g. \cite{cl95,smw97,wsk97,cl98,song9705} and 
references therein), and orbital structure (see e.g. \cite{song0012} 
and references therein) of the nucleon. In this model, the nucleon 
structure is determined by its zeroth order valence quark configuration 
and all possible quantum fluctuations of the valence quark into a recoil 
quark and a meson. It provides a natural and nonperturbative explanation 
of a nonzero (intrinsic) strange sea and negative strange sea polarization 
through the quark-kaon fluctuation (e.g. $u\to s+K^+$). The results 
agree quite well with the data obtained in deep-inelastic scattering
and other related experiments.  

Based on earlier encouraging result \cite{song9705}, we suggested 
an SU(4) version of quark-meson fluctuation model and predicted a 
nonzero intrinsic charm in the proton \cite{songictp98}. In the SU(4) 
version, charm or anti-charm quarks reside in the charmed mesons which 
are created by nonperturbative quantum quark-meson fluctuations such as 
$u\to c+\bar D^0$. These charm or anti-charm quarks are essentially 
$intrinsic$ components of the hadron. 

The main motivations of extending to SU(4) version are: 1) There is 
no physical reason to forbid the fluctuation from a light quark to a 
charmed quark and charmed meson. Hence it is interesting to investigate 
what are the changes of the SU(3) predictions after including the charm 
or anticharm contributions. 2) According to the symmetric GIM model 
\cite{gim70}, one should deal with the weak axial current in the 
framework of SU(4) symmetry. It implies that the charm quark should 
play some role in determining the spin, flavor and orbital structure 
of the nucleon. Therefore it is natural and reasonable to extend the
SU(3) version to SU(4) by including the charm and anticharm contributions. 

In this paper, we will discuss the effect of the IC contribution to 
the flavor and spin observables of the proton in the SU(4) quark meson 
fluctuation model with symmetry-breaking. The analytic and numerical 
results of the flavor and spin contents for each quark flavor $u$, $d$, 
$s$, and $c$ in the proton are presented. Predictions are compared with 
the existing data and other model results or analyses. The intrinsic charm 
contribution to the Ellis-Jaffe sum rule is also discussed. 
\par

\leftline{\bf 2. SU(4) model with symmetry breaking}
\par

In the framework of SU(4) quark model, there are sixteen pseudoscalar 
mesons, a 15-plet and a singlet. In this paper, the contribution of
the SU(4) singlet will be neglected. It is easy to show that the 
effective Lagrangian describing interaction between quarks and the 
mesons in the SU(4) case is
\be
{\it L}_I=g_{15}{\bar q}\pmatrix{{G}_u^0
&{\pi}^+ & {\sqrt{\epsilon}}K^+ & {\sqrt{\epsilon_c}}{\bar D}^0 \cr 
{\pi}^-& {G}_d^0 & {\sqrt{\epsilon}}K^0& {\sqrt{\epsilon_c}}D^-\cr
{\sqrt{\epsilon}}K^-& {\sqrt{\epsilon}}{\bar K}^0&{G}_s^0 &
{\sqrt{\epsilon_c}}{D}_s^-\cr 
{\sqrt{\epsilon_c}}{D}^0 & {\sqrt{\epsilon_c}}{D}^+
& {\sqrt{\epsilon_c}}{D}_s^+ &{G}_c^0 \cr }q, 
\label{(1)}
\ee 
where 
$D^+=(c\bar d)$, $D^-=(\bar cd)$, $D^0=(c\bar u)$, $\bar D^0=(\bar cu)$,
$D_s^+=(c\bar s)$, and $D_s^-=(\bar cs)$. The neutral charge components
${G}_{u(d)}^0$ and ${G}_{s,c}^0$ are defined as
\be
{G}_{u(d)}^0=+(-){\pi^0}/{\sqrt 2}+{\eta^0}\sqrt{\epsilon_{\eta}/6}+
{\eta'^0}\sqrt{\zeta'^2/48}-{\eta_c^0}\sqrt{\epsilon_c/16}
\label{(2)}
\ee
\be
{G}_s^0=-{\eta^0}\sqrt{2\epsilon_\eta/3}+{\eta'^0}\sqrt{\zeta'^2/48}
-{\eta_c^0}\sqrt{\epsilon_c/16}
\label{(3)}
\ee
\be
{G}_c^0=-{\eta'^0}\sqrt{3\zeta'^2/16}+{\eta_c^0}\sqrt{9\epsilon_c/16}
\lab{(4)}
\ee
with 
\be
\pi^0=(u\bar u-d\bar d)/\sqrt 2,~~\eta^0=(u\bar u+d\bar d-2s\bar s)\sqrt 6, 
\label{(5)}
\ee
\be
\eta'^0=(u\bar u+d\bar d+s\bar s)/\sqrt 3,~~\eta_c^0=(c\bar c).
\label{(6)}
\ee
Similar to the SU(3) case, we define $a\equiv |g_{15}|^2$, 
which denotes the transition probability of splitting $u(d)\to d(u)+
\pi^{+(-)}$. Hence $\epsilon a$, $\epsilon_{\eta} a$, $\zeta'^2 a$, and 
$\epsilon_c a$ denote the probabilities of splittings $u(d)\to s+K^{-(0)}$, 
$u(d)\to u(d)+\eta^0$, $u(d)\to u(d)+\eta'^0$ and $u(d)\to c+{\bar 
D}^{0}(D^-)$ respectively. If the splitting probability is dominated by 
the mass effects, we expect $0<\epsilon_{c} a<\zeta'^2 a<\epsilon 
a\simeq\epsilon_{\eta} a<a$, or 
\be
0~<~\epsilon_{c}~<~\zeta'^2~<~\epsilon~\simeq~\epsilon_{\eta}~<~1.
\label{(7)}
\ee
It implies that the probability of emitting heavier meson such as $D$ 
from light quarks is smaller than that of emitting the lighter mesons 
such as $K$, $\eta$, and $\eta'$, etc. Since the 
charm contributions only appeared in the observables which are explicitly 
related to the charm or anticharm variables, therefore we expect that 
the SU(4) model will not significantly change the SU(3) predictions on 
observables which are not explicitly charm-related. On the other hand, 
the SU(4) version leads to many $new$ predictions on observables which 
are directly related to the charm or anticharm. These predictions can be 
tested by the future experiments. Before going to further discussion, we 
would like to make a few remarks.
\begin{itemize}
\item{The quark contents of the $\eta^0$ and $\eta'^0$ shown in Eqs.(5) 
and (6) imply that they are actually $\eta_8$ and $\eta_1$. Hence the 
octet-singlet mixing is neglected in our description.}
\item{We have shown in \cite{song9705} that a better SU(3) description 
can be achieved by taking $\epsilon_\eta\simeq\epsilon$ and $\zeta'^2<<1$, 
i.e. the effective U(1)-breaking parameter is smaller than other
SU(3)-breaking parameters. This result was deduced from the data of 
$\bar d-\bar u$, which requires $\zeta'$ should be small and negative. 
More detail discussion can be found in section II and Fig.1 in 
\cite{song9705}. It can be shown that including the charm contributions 
will not change this conclusion. Therefore, we may choose $\zeta'^2=0$
as one possible option in the numerical calculation, i.e. neglect the 
singlet contribution from the beginning. In this case, the number of 
independent parameters would be only $three$: $a$, $\epsilon$, and 
$\epsilon_c$. Since our analytical formulae are not restricted to this 
approximation, we can also discuss $\zeta'^2\neq 0$ case (see discussion 
in section 4 below).}
\item{It should be noted that the definition of $G_{u(d)}^{0}$ in SU(4)
case is different from that in the SU(3) case. Besides an additional 
{\it charm} term, $-\eta_c^0\sqrt{\epsilon_c/16}$, the coefficient of 
$\eta^0$ term is changed. For $G_{s}^{0}$, even the coefficient of 
$\eta'^{0}$ term is also changed. Therefore, the SU(4) formalism cannot 
be reduced to SU(3) simply by taking $\epsilon_c\to 0$ only [see later 
discussion on Eqs. (15) and (16)]. The $G_c^{0}$ is completely new in 
the SU(4) version.} 
\item{As pointed out in the original chiral quark model
\cite{ehq92,cl95,smw97,wsk97,cl98,song9705}, the nucleon properties 
are defined in the scale range between $\Lambda_{\rm QCD}$ 
($\sim 0.2-0.3$ GeV) and $\Lambda_{\chi{\rm SB}}$ ($\sim 1$ GeV), 
where the spontaneous breaking of chiral symmetry leads to the 
existence of Goldstone bosons. In the quark meson fluctuation model 
(SU(3) in \cite{song0012} and SU(4) in this paper), we will assume 
all calculated quantities are also defined in the same scale range
[(0.2 GeV)$^2<\mu^2<$(1.0 GeV)$^2$]. At this $\mu^2$ range, the sea 
content should be dominated by the intrinsic component created by 
nonperturbative mechanism, quark-meson fluctuation, \cite{note1} 
and the extrinsic sea component is expected to be small.}
\item{We note that in our formalism, only the integrated flavor 
content [$q(Q^2)=\int_0^1dxq(x,Q^2)$] and helicity content [$\Delta 
q(Q^2)=\int_0^1dx\Delta q(x,Q^2)$] are discussed. To make brief of 
the formalism, we will omit the argument $Q^2$ everywhere unless 
specified otherwise.} 
\end{itemize}

In addition to the allowed fluctuations discussed in the SU(3) case, 
a valence quark is now allowed to split up or fluctuate to a recoil 
charm quark and a charmed meson. For example, a valence u-quark 
with spin-up, the allowed fluctuations are
\bea
u_{\up}&\to& d_{\dw}+\pi^+,~~u_{\up}\to s_{\dw}+K^+,~~
u_{\up}\to u_{\dw}+{G}_u^0,\label{(8)}
\\
u_{\up}&\to& c_{\dw}+{\bar D}^0,\label{(9)}
\\
u_{\up}&\to& u_{\up}.\label{(10)}
\eea

Similarly, one can list the allowed fluctuations for $u_{\dw}$,
$d_{\up}$, $d_{\dw}$, $s_{\up}$, and $s_{\dw}$. Since we are only 
interested in the spin-flavor structure of the nucleon (or other
noncharmed hadrons), which does not have valence charm quark, hence 
the fluctuations from a valence charm quark will not be discussed. 
 
The spin-up and spin-down quark or antiquark contents in the proton,
up to first order of the quantum fluctuation, can be written as
(a detail SU(3) version see e.g. \cite{song0012}), 

\be
n_p(q'_{\up,\dw}, {\rm or}\ {\bar q'}_{\up,\dw}) 
=\sum\limits_{q=u,d}\sum\limits_{h=\up,\dw}
n^{(0)}_p(q_h)P_{q_h}(q'_{\up,\dw}, {\rm or}\ {\bar q'}_{\up,\dw}),
\label{(11)}
\ee
where $q'=u,d,s,c$ and $n_p^{(0)}(q_{\up,\dw})$ are determined by the 
zeroth order, i.e. naive quark model (NQM), valence quark wave function 
of the proton and
\bea
n^{(0)}_p(u_{\up})&=&{5/3},~~~~n^{(0)}_p(u_{\dw})={1/3},\non \\
& & \label{(12)}\\
n^{(0)}_p(d_{\up})&=&{1/3},~~~~n^{(0)}_p(d_{\dw})={2/3}.\non
\eea
In Eq.(11), $P_{q_{\up,\dw}}(q'_{\up,\dw})$ and $P_{q_{\up,\dw}}({\bar
q}'_{\up,\dw})$ are probabilities of finding a quark $q'_{\up,\dw}$
or an antiquark $\bar q'_{\up,\dw}$ arise from all quantum fluctuations 
of a valence quark $q_{\up,\dw}$. The probabilities
$P_{q_{\up,\dw}}(q'_{\up,\dw})$ and $P_{q_{\up,\dw}}({\bar q}'_{\up,\dw})$ 
can be obtained from the effective Lagrangian (1) and have been listed 
in Table I, where only $P_{q_{\up}}(q'_{\up,\dw})$ and 
$P_{q_{\up}}({\bar q}'_{\up,\dw})$ are shown. Those arise from 
$q_{\dw}$ can be obtained by using  
$P_{q_{\dw}}(q'_{\up,\dw})=P_{q_{\up}}(q'_{\dw,\up})$ and
$P_{q_{\dw}}({\bar q}'_{\up,\dw})=P_{q_{\up}}({\bar q}'_{\dw,\up})$.
The notations appeared in Table I are defined as 
\bea
f&\equiv& {1/2}+\epsilon_{\eta}/6+\zeta'^2/48+\epsilon_c/16,\non \\
& & \label{(13)}\\
f_s&\equiv& 2\epsilon_{\eta}/3+\zeta'^2/48+\epsilon_c/16,\non
\eea
and
\bea
\tilde A\equiv {1/2}&-&\sqrt{\epsilon_\eta}/6
-\zeta'/12,~~~\tilde B\equiv -\sqrt{\epsilon_\eta}/3
+\zeta'/12,\non \\
&&\label{(14)}\\
\tilde C&\equiv& 2\sqrt{\epsilon_\eta}/3+\zeta'/12,~~~~
\tilde D\equiv \sqrt{\epsilon_c}/4.\non
\eea

Analogous to the SU(3) case, the special combinations $\tilde A$, 
$\tilde B$, $\tilde C$, and $\tilde D$ stem from the definitions 
(2)-(4), which show the quark and antiquark contents in the neutral 
bosons ${G}_{u,d,s,c}^0$. The numbers $fa$ 
and $f_sa$ stand for the probabilities of the quark splitting 
$u_{\up}(d_{\up})\to u_{\dw}(d_{\dw})+{G}_{u(d)}^0$ and 
$s_{\up}\to s_{\dw}+{G}_s^0$ respectively.
 
To reduce to the SU(3) version, one should take the limit 
$\epsilon_c\to 0$ and change $\zeta'$ to $4\zeta'$ simultaneously. 
This is because we have different neutral combinations $G_{u,d,s}^{0}$ 
from the SU(3) ones. This difference shows that the $charm$ 
contribution is not only presented in the $charm$ term which is 
proportional to $\epsilon_c$, but also affected by the $\zeta'$ term. 
Taking $\epsilon_c\to 0$ and $\zeta'\to 4\zeta'$, one can see from Eq. 
(13) that the $f$ and $f_s$ indeed reduce to the corresponding 
quantities in the SU(3) case,
\bea
(f)_{SU(3)}&=& {1/2}+\epsilon_{\eta}/6+\zeta'^2/3,\non \\
& & \label{(15)}\\
(f_s)_{SU(3)}&=& 2\epsilon_\eta/3+\zeta'^2/3,
\non
\eea
which is exactly Eq.(6b) in \cite{song0012}. Taking $\epsilon_c\to 0$ 
and $\zeta'\to 4\zeta'$, one has $\tilde D\to 0$ and 
\be
{\tilde A}\to A_{SU(3)}/3,~~
{\tilde B}\to B_{SU(3)}/3,~~
{\tilde C}\to C_{SU(3)}/3,
\lab{(16)}
\ee
where $A_{SU(3)}$, $B_{SU(3)}$, and $C_{SU(3)}$ are the same as 
$A$, $B$, and $C$ in Eq. (6a) in \cite{song0012}. 

If Eq.(12) is replaced by $n^{(0)}_n(d_{\up})={5/3}$, 
$n^{(0)}_n(d_{\dw})={1/3}$, $n^{(0)}_n(u_{\up})={1/3}$, and 
$n^{(0)}_n(u_{\dw})={2/3}$, from (11), we will obtain all flavor and 
spin contents in the neutron.
\par

\leftline{\bf 3. Flavor and spin contents in the proton}
\par

We note that the quark {\it flips} its spin in the quark splitting 
processes $q_{\up,(\dw)}\to q_{\dw,(\up)}$+meson, i.e. processes in 
(8) and (9), but not in the (10) ($no$ $splitting$). The quark helicity 
non-flip contribution in splitting processes is entirely neglected. 
This is a basic assumption in the model and seems to be consistent 
with the picture given by the instanton model. 

\leftline{\quad \bf 3.1. Quark flavor content}
\par

Using (11), (12) and the probabilities $P_{q_{\up,\dw}}(q'_{\up,\dw})$ 
and $P_{q_{\up,\dw}}({\bar q}'_{\up,\dw})$ listed in Table I, one  
obtains all quark and antiquark flavor contents in the proton, 
\be
u=2+\bar u,~d=1+\bar d,~s=0+\bar s,~c=0+\bar c,~
\label{(17)}
\ee
where 
\be
\bar u=a[1+\tilde A^2+2(1-\tilde A)^2],~
\bar d=a[2(1+\tilde A^2)+(1-\tilde A)^2],~
\label{(18)}
\ee
\be
\bar s=3a[\epsilon+\tilde B^2],~~~\bar c=3a[\epsilon_c+\tilde D^2].
\label{(19)}
\ee
From (18), one obtains
\be
{\bar u}/{\bar d}=1-6\tilde A/[(3\tilde A-1)^2+8],
\label{(20)}
\ee
and
\be
\bar d-\bar u=2a\tilde A.
\label{(21)}
\ee
Similarly, one can obtain $2\bar c/(\bar u+\bar d)$, $2\bar c/(u+d)$, 
$2\bar c/\sum(q+\bar q)$ and other flavor observables related to charm 
quark and cahrm antiquark. We note that in the SU(4) case, one has
\be
\sum\limits_{q,\bar q}(q+\bar q)=3(1+2a\xi_1),
\label{(22)}
\ee
where $\xi_1\equiv 1+\epsilon+\epsilon_c+f$. It is easy to verify that 
in the limit $\epsilon_c\to 0$, and using Eq.(16), all SU(4) results 
reduce to those given in the SU(3) case. Two remarks on (20) and (21) 
are in order.
\begin{itemize}
\item{Defining the ratio, $r\equiv\bar u/\bar d$, we obtain from Eq. 
(20) that $\tilde A=2-r\pm\sqrt{14r-8r^2-5}$. Since $\tilde A$ must be 
a $real$ number due to (21), hence $14r-8r^2-5$ must be positive or 
zero, and the ratio $r$ satisfies 
\be
{1/2}~\leq~ \bar u/\bar d~ \leq ~{5/4}, 
\lab{(23)}
\ee
which seems to be consistent with the experimental data shown in Table 
III.}
\item{Since $|\zeta'/2|\leq \sqrt{\epsilon_\eta}$ in our 
approximation, from (14), we have $\tilde A\leq 1/2$ (note that 
$\zeta'$ is $negative$) and 
\be
\bar d-\bar u\leq a.
\lab{(24)}
\ee
It gives a lower bound of the parameter $a$ $-$ the probability of quark 
splitting $q\to q'+\pi$.}
\end{itemize}

\leftline{\quad \bf 3.2. Quark helicity content}
\par

Similar to the SU(3) version, we have
\be
\Delta q=\sum\limits_{q'=u,d} [n^{(0)}({q'}_{\up})-n^{(0)}({q'}_{\dw})]
[P_{{q'}_{\up}}(q_{\up})-P_{{q'}_{\up}}(q_{\dw})].
\lab{(25)}
\ee
Using Eqs. (12), (25), and Table I, we obtain
\be
\Delta u=4[1-a(\epsilon+\epsilon_c+2f)]/3-a,
\lab{(26)}
\ee
\be
\Delta d=-[1-a(\epsilon+\epsilon_c+2f)]/3-a,
\lab{(27)}
\ee
\be
\Delta s=-a\epsilon,
\lab{(28)}
\ee
\be
\Delta c=-a\epsilon_c, 
\lab{(29)}
\ee
\be
\Delta\Sigma\equiv\sum\limits_{q=u,d,s,c}\Delta
q=1-2a(1+\epsilon+\epsilon_c+f),
\label{(30)}
\ee
or
\be
\Delta\Sigma/2={1/2}-a\xi_1,
\label{(31)}
\ee
where $\xi_1$ is defined in (22), and
\be
\Delta{\bar q}=0,~~~({\bar q}={\bar u},{\bar d},{\bar s},{\bar 
c}). 
\lab{(32)}
\ee
Before going to the numerical calculation, we would like to make several 
remarks on some results which are basically independent of parameters.
\begin{itemize}
\item{Comparing with the SU(3) case, a new $\epsilon_c$ term, which 
presents the intrinsic charm contribution, appeared in $\Delta u$, 
$\Delta d$, $\Delta c$, and $\Delta\Sigma$, but not in $\Delta s$.
This is because there is no process which can mix the $strange$ 
helicity and $charm$ helicity contributions. $\Delta s$ can only
come from the processes like $u_{\up}\to s_{\dw}+K^+$, while
$\Delta c$ comes only from the processes like $u_{\up}\to c_{\dw}+{\bar 
D}^0$. Although there are $s$, $\bar s$ (or $c$, $\bar c$) in all 
$neutral$ bosons ${G}_{u,d,s,c}^0$, they give $no$ contributions
to $\Delta s$ (or $\Delta c$) due to $s_\up$ and $s_\dw$ (or $c_\up$
and $c_\dw$) appeared with $equal$ $probability$ in these neutral 
bosons.}
\item{The charm quark helicity $\Delta c$, (29), is {\it nonzero} as 
far as $\epsilon_c$ is nonzero. Analogous to the strange quark helicity,
$\Delta c$ is definitely {\it negative}, but the size of the intrinsic 
charm helicity depends on the parameter $\epsilon_c$ and the splitting 
probability $a$.}
\item{The physical meaning of (31) is that the total $loss$ of the
quark helicity arises from $four$ splitting processes with quark
spin-flip, $three$ in (8) and $a$ $new$ splitting in (9). Comparing 
with the SU(3) case, where $\Delta\Sigma/2=1/2-a(1+\epsilon+f)$, we 
now have an additional reduction, $-a\epsilon_c$, of the total quark 
spin due to the splitting related to the charm.} 
\item{In the splitting process $u_{\up(\dw)}~\to~c_{\dw(\up)}+\bar D^0$, 
the anticharm resides only in the charmed meson, e.g. $\bar D^0(\bar 
c,u)$. The probabilities of finding $\bar c_{\up}$ and $\bar c_{\dw}$ 
are equal in the spinless charmed meson. Therefore $\Delta\bar c=0$. 
Similar discussion in the SU(3) case \cite{smw97} has led to $\Delta
\bar q=0$ for $\bar q=\bar u,\bar d,\bar s$. The result (32) shows 
that the helicities of the sea quark and antiquark are not equal, 
$\Delta q_{sea}\neq\Delta\bar q$. This is different from the usual
gluon splitting ($g\to q+\bar q$) model and $gg\to q+\bar q$ model
\cite{stan81}. 
In the gluon splitting and gluon fusion models, the sea quark and 
antiquark with the same flavor are perturbatively or nonperturbatively 
created as a {\it pair} from the gluon or gluons and $\Delta 
q_{sea}=\Delta\bar q$. The DIS data \cite{adeva96} seems to support 
the prediction $\Delta\bar q\simeq 0$ but with large errors. }
\item{From (19) and (29), using ${\tilde D}^2={\epsilon_c}/16$ in
(14), one can see that the ratio 
\be
{{\Delta c}/{c}}=-{{16/51}}
\label{(33)}
\ee
is a constant {\it independent} of any splitting parameters. 
This is a special prediction for the charm flavor in the SU(4) quark 
meson model. }
\item{For the strange flavor, from $\tilde B$ in (14) and $\bar s$ in
(19), one has 
$s=\bar s\simeq\epsilon a(10/3)[1+|\zeta'|/(20\sqrt\epsilon)]$ due to 
$\zeta'^2<<1$. In the limit $\zeta'^2\to 0$, we have 
\be
{{\Delta s}/{s}}=-{{3/10}}.
\label{(34)}
\ee
Hence $\Delta s/s$ is also a constant in this limit.}
\item{From (33) and (32), one obtains the ratio of the charm with
positive helicity to that with negative helicity is also a
constant,
\be
{{c_\up/c_\dw}}={{35/67}}\simeq 0.522.
\label{(35)}
\ee
It shows that more $c_{\dw}$ is created than $c_{\up}$ in the 
splitting processes $u_{\up(\dw)}~\to~c_{\dw(\up)}+\bar D^0$ and 
$d_{\up(\dw)}~\to~c_{\dw(\up)}+D^-$. This is because the total 
contribution from $u_{\up}$ and $d_{\up}$ splittings is larger than 
that from $u_{\dw}$ and $d_{\dw}$ splittings.
Similar situation occurs for the strange quarks. From (34)
and (32), we have, for $\zeta'=0$,
\be
{{s_\up/s_\dw}}={{7/13}}\simeq 0.538.
\label{(36)}
\ee}
\item{For the $u$-flavor and $d$-flavor, we do not have similar exact
relations like (33) and (34). This is because the quantities 
$u_\up-u_\dw$ and $d_\up-d_\dw$ $depend$ $on$ $\epsilon_c$, while
$u_\up+u_\dw$ and $d_\up+d_\dw$ do not. Hence ${{\Delta u}/{u}}$ 
and ${{\Delta d}/{d}}$ depend on all three splitting parameters.
The numerical calculation (see next section) shows that 
\be
{{\Delta u}/{u}}\simeq 0.383,~~~{{\Delta d}/{d}}\simeq -0.287.\lab{(37)}
\ee
From (37) and (32), we obtain 
\be
{{d_\up/d_\dw}}\simeq 0.554,
\label{(38)}
\ee
and 
\be
{{u_\up/u_\dw}}\simeq 2.241.
\label{(39)} 
\ee
Comparing with the zeroth approximation (NQM), where 
${{d_\up}/{d_\dw}}=1/2$ and ${{u_\up}/{u_\dw}}=5/1$, hence the 
quark-meson splittings lead to a small enhancement for 
$d_\up/d_\dw$, but a significant reduction for $u_\up/u_\dw$.} 
\item{The ratios of the flavor or spin contents predicted in this paper 
are also defined at $Q^2=\mu^2$ scale, where the sea (light or heavy) 
quark contents are `intrinsic' and determined by nonperturbative 
mechanism. To compare the predictions with data at higher $Q^2$ range, 
one needs to use perturbative QCD and treat these ratios (at $Q^2=
\mu^2$) as inputs, i.e. boundary conditions, in the QCD evolution 
equations. If the numerator and denominator of the ratio have the 
same or almost the same factor of $Q^2$ dependence, then this ratio 
will not sensitive to the change of $Q^2$.   }
\end{itemize}

\leftline{\quad \bf 3.3. Ellis-Jaffe sum rule}
\par

In the framework of SU(4) parton model, the first moment of the spin
structure function $g_1^p(x,Q^2)$ in the proton is
\bea
\Gamma_1^p&\equiv&\int_0^1g_1^p(x,Q^2)dx
\non \\
&=&(4\Delta u+\Delta d+\Delta s+4\Delta c)/18,
\label{(40)}
\eea
which can be rewritten as
\bea
\Gamma_1^p&\equiv&\int_0^1g_1^p(x,Q^2)dx
\non\\
&=&(3a_3+a_8-a_{15}+{5}a_0)/36
\label{(41)}
\eea
where the notations 
\be
a_3=\Delta u-\Delta d,~~
a_8=\Delta u+\Delta d-2\Delta s
\label{(42)}
\ee
\be
a_{15}=\Delta u+\Delta d+\Delta s-3\Delta c
\label{(43)}
\ee
\be
a_0=\Delta u+\Delta d+\Delta s+\Delta c
\label{(44)}
\ee
have been introduced. Using exchange $\Delta u\leftrightarrow \Delta d$,
or $a_3\leftrightarrow -a_3$ in (41), we can obtain $\Gamma_1^n$ for the
neutron.
\par

\leftline{\bf 4. Numerical results and discussion.}
\par

Since the probability of charm-related splitting (9) is much smaller
than $u$-,$d$-,$s$-related splittings (8), we expect that the values 
of parameters $a$ and $\epsilon$ in SU(4) should be very close to 
those used in the SU(3) version. We choose $a=0.143$, $\epsilon=0.454$ 
[note that $a=0.145$, $\epsilon=0.460$ in SU(3)], and leave $\epsilon_c$ 
as a $variable$, then express the quark flavor and helicity contents 
as functions of $\epsilon_c$. To determine the range of $\epsilon_c$, 
we plot $a_3$ as the function of $\epsilon_c$ in Fig.1. Using the most 
precise data from the neutron $\beta$-decay \cite{pdg00}, 
$a_3=(G_A/G_V)_{n\to p}=1.2670\pm 0.0035$, we obtain 
$\epsilon_c\simeq 0.06\pm 0.02$. To include other possible theoretical 
uncertainties arise from the model approximations, however, we prefer 
to introduce a larger uncertainty and take  
\be
\epsilon_c\simeq 0.06\pm 0.04.
\label{(45)}
\ee 

In addition to the three-parameter set $\{a, \epsilon, \epsilon_c\}$
[SU$^{(a)}$(4)], we also consider the four-parameter set:
$\{a, \epsilon, \zeta', \epsilon_c\}$ [SU$^{(b)}$(4)] to show the 
$\eta'^0$ effect. The parameter sets used in this paper and in previous 
SU(3) version are listed in Table II.

Using the parameter sets SU$^{(a)}$(4) and SU$^{(b)}$(4), the flavor 
and spin observables in the proton are calculated and listed in Tables 
III and IV respectively. For comparison, we also list the existing data, 
the results from previous SU(3) description \cite{song9705,song0012}, 
the naive quark model (NQM) prediction, and results given by other 
models or analyses. Including the QCD radiative corrections, the first 
moments of the spin structure functions $g_1^{p}$ and $g_1^{n}$ are 
also shown in Table IV. We also calculated two reduced matrix elements 
$F$ and $D$, and the ratios $G_A/G_V$ of the hyperon beta decays. The 
results are listed in Table V. One can see that both SU$^{(a)}$(4) and 
SU$^{(b)}$(4) versions satisfactorily describe almost all the existing 
data. Furthermore, they give many new predictions on charm-related 
observables, which are in the bold type shown in Tables III and IV 
(for simplicity, we use $\epsilon_c=0.06$ in SU$^{(b)}$(4) without 
the uncertainty). A few main predictions selected from SU$^{(a)}$(4) 
results are listed in the following:
\be
2\bar c/\sum(q+\bar q)\simeq 0.011\pm 0.008
\label{(46)}
\ee
measures the size of the intrinsic charm in the proton, 
\be
\Delta c\simeq -0.009\pm 0.006
\label{(47)}
\ee
is the charm helicity, and
\be
{{\Delta c/\Delta\Sigma}}\simeq -(0.021\pm 0.014) 
\label{(48)}
\ee
is the fraction of total quark helicity carried by the charmed quark. 
Several remarks are in order:
\begin{itemize}
\item{Comparing the SU$^{(a)}$(4) with SU$^{(b)}$(4) predictions, 
one can see that there are 10-20$\%$ differences between them only
for the light-flavor contents, e.g. $\bar d-\bar u$ (0.111 -$>$ 0.141), 
$\bar d/\bar u$ (0.71 -$>$ 0.64), $2\bar s/(\bar u+\bar d)$ 
(0.66 -$>$ 0.75), and $2\bar s/(u+d)$ (0.118 -$>$ 0.133) [we note that
both predictions are consistent with the existing data within errors], 
and all other predictions are almost no changes. It shows that the 
$\eta'^0$ has minor or no effect on the spin and charm-related 
observables in the model. This is because first the
$\eta'^0$ does not have $c, \bar c$ components and second the 
$q_{\up}$ and $q_{\dw}$ (or $\bar q_{\up}$ and $\bar q_{\dw}$), 
$q=u,d,s$, appeared with equal probability in $\eta'^0$.  
It should be noted that the approximation $\zeta'^2\simeq 0$
does not mean the probability of fluctuation $q\to q'+\eta'^0$ 
($\zeta'^2a$) is smaller than that of $q\to q'+\bar D^0$ 
($\epsilon_c a$). Actually, we have $\epsilon_c<\zeta'^2$.
Hence we use the approximation $\zeta'^2=0$ in SU$^{(a)}$(4) version
only for the practical reason explained here.} 
\item{The probability $a$ of the fluctuation $u\to d+\pi^+$ has been 
estimated in chiral field theory (see e.g. \cite{ehq92,song9705})
\bea
a_{u\to d+\pi^+}&=&\int_0^1dz\Theta(\Lambda^2-\tau(z))P_{u\to 
d+\pi^+}(z),
\label{(49)}
\eea
where
\bea
P_{u\to d+\pi^+}(z)&=& 
{{g_A^2(m_u+m_d)^2}\over {32\pi^2{f_\pi}^2}}z\cdot \non\\
&&\int_{-\Lambda^2}^{-\tau(z)}dt{{t-(m_u+m_d)^2}\over 
{(t-m_{\pi}^2)^2}},
\label{(50)}
\eea
and $-\tau(z)=m_u^2z-m_{d}^2z/(1-z)$. In Eq.(50),
$g_A\simeq 0.75$ is the dimensionless axial-vector coupling,
$f_\pi\simeq 0.093$ GeV the pion decay constant, $m_u$ ($m_d$) the 
constituent mass of the $u$ ($d$) quark, $\Lambda$ the ultraviolet 
cutoff, and $m_\pi$ is the pion mass. For $\Lambda\simeq 2.4$ GeV, 
one obtains $a\simeq 0.142$. For the strange quark fluctuation, 
$u\to s+K^+$, similar calculation with some approximation leads to 
the probability $\epsilon a\simeq 0.062$, which gives $\epsilon\simeq 
0.44$. We assume the same formula can be used for the charm fluctuation, 
$u\to c+\bar D^0$, and obtain $\epsilon_c a\simeq 0.004$, which gives
$\epsilon_c\simeq 0.03$, which is consistent with $\epsilon_c\simeq 
0.06\pm 0.04$ given in Eq.(45). Physically, the probability of splitting 
to the heavier mesons should be less than of splitting to the lighter 
ones. Hence the above estimation is quite reasonable.}
\item{The theoretical uncertainties shown in (46)-(48) and 
in Tables III, IV and V arise only from the uncertainty of $\epsilon_c$ 
in (45). If the observable does not depend on $\epsilon_c$, such as 
$\bar d-\bar u$, $\bar d/\bar u$, $2\bar s/(\bar u+\bar d)$, etc. 
(these quantities depend only on $\tilde A$, $\tilde B$, and $\tilde C$, 
and not on $\tilde D$, i.e. independent of $\epsilon_c$), there is no 
uncertainty for them. This has been shown in Table III. Two special 
quantities $\Delta c/c$ and $\Delta s/s$ are also independent of 
$\epsilon_c$ as shown in Table IV. We put a star (*) mark on some `data' 
in Tables III and IV to denote they are model predictions or from 
theoretical analyses.}
\item{The SU(4) quark-meson model predicts an intrinsic charm component 
of the nucleon, ${2\bar c}/\sum(q+\bar q)\simeq 1\%$, which agrees with 
the predictions given in \cite{dg77,nnnt96,gol00} and is also close to the 
those given in \cite{hk94,pst99} and \cite{bs91,smt99}. We note that the 
IC component is almost one order of magnitude smaller than the intrinsic
strange component ${2\bar s}/\sum(q+\bar q)$. }
\item{Using the approach given in a previous work (see Eq. (3.6) in 
\cite{song9705}), we can show that 
\bea
2&\int_0^1&dxx\bar c(x)/\sum\int_0^1dxx[q(x)+\bar q(x)]\non\\
&<& 2\bar c/\sum(q+\bar q),
\label{(51)}
\eea
where, as defined in this paper, $\bar c\equiv \int_0^1dx\bar c(x)$ 
and $\sum(q+\bar q)\equiv\sum\int_0^1dx[q(x)+\bar q(x)]$.
The l.h.s. of Eq. (51) is the fraction of the total quark momentum 
carried by the charm and anticharm quarks. The prediction (46) 
implies that this fraction is less than $1\%$. Assuming the quark and 
antiquark share about one half of the nucleon momentum, then the charm 
and anticharm carry about $0.5\%$ of the nucleon momentum or less.} 
\item{From Table III, we have
\bea
u+\bar u~&:&~d+\bar d~:~s+\bar s~:~c+\bar c\nonumber
\\
\simeq 0.53~&:&~0.37~:~0.09~:0.01.
\label{(52)}
\eea
If we assume the quarks carry about $55\%$ of the nucleon momentum,
Eq. (52) implies that the fractions of the nucleon momentum carried 
by $u$-, $d$-, $s$-, and $c$-flavors are approximately $29.2\%$, 
$20.3\%$, $4.9\%$, and $0.6\%$ respectively. They may compare with 
$31.4\%$, $17.8\%$, $4.3\%$, and $1.2\%$ given by the DIS data at 
$Q^2$=20 GeV$^2$, where the gluons carry about $45\%$ of the nucleon 
momentum.}
\item{The prediction of intrinsic charm polarization, $\Delta c\simeq
-0.009\pm 0.006$ is close to the result $\Delta c=-0.020\pm 
0.005$ given in the instanton model \cite{amt98}. This might 
be related to the strong suppression of non-spinflip contribution in 
both models. Our $\Delta c$ in (47) is smaller in magnitude than 
that given in \cite{blos98,hz97} ($\simeq -0.3$). However, the size 
of $\Delta c$ given in \cite{pst99} ($\simeq -5\cdot 10^{-4}$) is even 
smaller. Hence further investigation in this quantity is needed.} 
\item{The ratio $\Delta c/\Delta\Sigma$ as the function of $\epsilon_c$ 
is plotted in Fig.2. Taking $\epsilon_c\simeq 0.06$, one has
$\Delta c/\Delta\Sigma\simeq -0.021$. This is consistent with the 
prediction given in \cite{amt98}, but smaller than that given in 
\cite{blos98}. Combining with the fractions of the light quark helicities, 
we have
\bea
{{\Delta u/\Delta\Sigma}}\simeq 2.171,~~ 
{{\Delta d/\Delta\Sigma}}\simeq -0.988,\non\\
\label{(53)}\\
{{\Delta s/\Delta\Sigma}}\simeq -0.162,~~ 
{{\Delta c/\Delta\Sigma}}\simeq -0.021, \non
\eea
one can see that the $u$-quark helicity is $positive$ ($parallel$ to the 
nucleon spin) and about two times larger than the total quark helicity 
$\Delta\Sigma$. The $d$-, $s$-, and $c$-helicities, however, are all
$negative$ ($antiparallel$ to the nucleon spin), and their sizes are
decrease as 
\be
|\Delta d|~:~|\Delta s|~:~|\Delta c|~\simeq~1~:~10^{-1}~:~10^{-2}.
\label{(54)}
\ee
Compare to the intrinsic strange helicity
$\Delta s$, the intrinsic charm helicity is one order of magnitude 
smaller.}
\end{itemize}

Since the hyperon $\beta$-decay data are measured at low $Q^2$ and
and our model predictions are defined at the scale (0.2 GeV)$^2<\mu^2 
<$(1.0 GeV)$^2$, hence we may compare them with less ambiguity. 
However, many data listed in Tables III and IV are coming from the 
DIS measurements at higher $Q^2$ range. To make a meaningful 
comparison of model predictions with these data, we have to discuss 
the $Q^2$ dependence of these observables. For spin observable, as 
we mentioned in section III of \cite{song9705} that the model 
predictions, e.g. $\Delta u$, $\Delta d$, $\Delta s$, etc. are 
compared with the (factorization) scheme-independent DIS observables 
$a_q(Q^2)\equiv\Delta q-[\alpha_s(Q^2)/2\pi]\Delta G(Q^2)$ $(q=u,d,s)$ 
at the same $Q^2$ scale [i.e. (0.2 GeV)$^2< \mu^2 <$ (1.0 GeV)$^2$], 
where  $\Delta G(Q^2)$ is the helicity of the gluon, and the axial 
charge $a_q(Q^2)$ is defined in the Adler-Bardeen scheme. Although 
$a_q(Q^2)$ is independent of $Q^2$ at the leading order and changes 
very slowly with $Q^2$ at NLO, we still need to assume the perturbative 
QCD can be used down to the scale $\mu^2$. The perturbative QCD 
evolution approach has been successfully used down to $Q^2\simeq 0.23$ 
GeV$^2$ \cite{grv95}, it is not clear, however, if the approach still 
hold below this $Q^2$. 

Finally, in the quark meson fluctuation model, it is possible to 
include the contributions come from quark splittings to the vector 
mesons such as $K^*$, $\rho$, etc. [$1^-$ nonet in SU(3) and $1^-$ 
15-plet in SU(4)]. However, it will change the formalism completely 
and is far beyond the goal of this paper. For simplicity and 
consistency, we only discuss the contributions of 0$^-$ pseudoscalar
mesons at this moment and defer the discussion of possible vector meson 
fluctuations to a later time. 
 
In summary, we have calculated the intrinsic charm contribution in the  
SU(4) quark meson model with symmetry breaking. Despite the approximations
and possible theoretical uncertainties, the overall agreement between the 
predictions and the existing data seems to be quite satisfactory 
considering the model is $simple$ and has only $a$ $few$ parameters. The 
model also leads to many $new$ predictions on observables explicitly 
related to the charm or anticharm. These observables are $zero$, e.g. 
$2\bar c/\sum(q+\bar q)$, $\Delta c$, etc. or $indefinite$, e.g. 
$\Delta c/c$ and $c_\up/c_\dw$, in the SU(3) description. We hope that 
these predictions can be tested by the analyses of the DIS data on 
polarized and unpolarized charm productions in the near future.
\par

\acknowledgments
This work was supported in part by the U.S. DOE Grant, the Institute 
of Nuclear and Particle Physics, Department of Physics, University of 
Virginia, and the Commonwealth of Virginia.
\par


\widetext
\begin{table}
\caption{The probabilities $P_{q_{\up}}(q'_{\up,\dw},\bar q'_{\up,\dw})$
and $P_{q_{\up}}(q'_{\up,\dw},{\bar q}'_{\up,\dw})$.} 
\begin{tabular}{|c|c|c|c|}
\hline 
$q'$ &$P_{u_{\up}}(q'_{\up,\dw})$ & $P_{d_{\up}}(q'_{\up,\dw})$ 
&$P_{s_{\up}}(q'_{\up,\dw})$ 
\\ 
\hline 
$u_{\up}$ & $1-(1+\epsilon+\epsilon_c+2f)a/2+
(1-\tilde A)^2a/2$ & $\tilde A^2a/2$ 
&$\tilde B^2a/2$
\\
$u_{\dw}$ & $(1+\epsilon+\epsilon_c+2f)a/2+
(1-\tilde A)^2a/2$ & $a+\tilde A^2a/2$ 
&$\epsilon a+\tilde B^2a/2$ 
\\
$d_{\up}$ & $\tilde A^2a/2$ &$1-(1+\epsilon+\epsilon_c+2f)a/2
+(1-\tilde A)^2a/2$ 
&$\tilde B^2a/2$ 
\\
$d_{\dw}$ & $a+\tilde A^2a/2$ &
$(1+\epsilon+\epsilon_c+2f)a/2+(1-\tilde A)^2a/2$ 
& $\epsilon a+\tilde B^2a/2$ 
\\
$s_{\up}$ & $\tilde B^2a/2$ &$\tilde B^2a/2$ 
&$1-(\epsilon+f_s+\epsilon_c/2)a+\tilde C^2a/2$ 
\\
$s_{\dw}$ & $\epsilon a+\tilde B^2a/2$ & $\epsilon a+\tilde B^2a/2$ 
&$(\epsilon+f_s+\epsilon_c/2)a+\tilde C^2a/2$ 
\\
$c_{\up}$ & $\tilde D^2a/2$ &$\tilde D^2a/2$ &$\tilde D^2a/2$ 
\\
$c_{\dw}$ & $\epsilon_c a+\tilde D^2a/2$ & $\epsilon_c a+\tilde D^2a/2$ 
&$\epsilon_c a+\tilde D^2a/2$ 
\\
\hline
${\bar u}_{\up,\dw}$ &$(1-\tilde A)^2a/2$ & $(1+\tilde A^2)a/2$ 
&$(\epsilon+\tilde B^2)a/2$ 
\\
${\bar d}_{\up,\dw}$ &$(1+\tilde A^2)a/2$&
$(1-\tilde A)^2a/2$ 
&$(\epsilon+\tilde B^2)a/2$ 
\\
${\bar s}_{\up,\dw}$ &$(\epsilon+\tilde B^2)a/2$&
$(\epsilon+\tilde B^2)a/2$
& $\tilde C^2a/18$ 
\\
${\bar c}_{\up,\dw}$ &$(\epsilon_c+\tilde D^2)a/2$&
$(\epsilon_c+\tilde D^2)a/2$
&$(\epsilon_c+\tilde D^2)a/2$
\\
\hline
\end{tabular}
\end{table}

\widetext
\begin{table}[h]
\caption{Parameter sets for different symmetry-breaking schemes}
\begin{tabular}{|c|c|c|c|c|} \hline
Model & $a$ & $\epsilon$ &  ${\zeta'^2}_{SU(3)}$ & $\epsilon_c$ \\
\hline
SU$^{(a)}$(4)        & 0.143 & 0.454 & $-$ & $0.06\pm 0.04$ \\
\hline 
SU$^{(b)}$(4)        & 0.143 & 0.454 & 0.1    & 0.06\\
\hline 
SU(3)\cite{song0012} & 0.145 & 0.460& 0.1 &  $-$        \\
\hline 
\end{tabular}
\end{table}

\widetext
\begin{table}
\nopagebreak
\caption{Quark Flavor Observables.}
\begin{tabular}{|c|c|c|c|c|c|} 
\hline
Quantity & Data & SU$^{(a)}$(4)&SU$^{(b)}$(4)& SU(3)\cite{song0012}&NQM\\
\hline 
$\bar d-\bar u$ & $0.110\pm 0.018$\cite{peng98}&0.111&0.141&0.143&0  \\
                & $0.147\pm 0.039$\cite{am94} &       &  &  &\\
\hline 
${{\bar u}/{\bar d}}$ &$[{{\bar u(x)/\bar d(x)}}]_{0.1<x<0.2}
=0.67\pm 0.06$\cite{peng98} & 0.71 &  0.64& 0.64&$-$\\
&$[{{\bar u(x)/\bar d(x)}}]_{x=0.18}=0.51\pm 0.06$\cite{na51} & 
& &&\\ 
\hline
${{2\bar s}/{(\bar u+\bar d)}}$ & ${{<2x\bar s(x)>}/{<x(\bar
u(x)+\bar d(x))>}}$ & 0.66& 0.75& 0.76&$-$\\
&$=0.477\pm 0.051$\cite{ba95}& &&&\\
\hline
${{2\bar c}/{(\bar u+\bar d)}}$ & $-$ & $\bf 0.083\pm 0.055$&0.085 &0 
&$-$\\
\hline
${{2\bar s}/{(u+d)}}$ & $<2x\bar s(x)>/<x(u(x)+d(x))>$&0.118 &0.133& 0.136&0\\
&$=0.099\pm 0.009$\cite{ba95}&&&&\\
\hline
${{2\bar c}/{(u+d)}}$ & $-$ & $\bf 0.015\pm 0.010$&0.015 &0&0\\
\hline
$(u+\bar u)/\sum(q+\bar q)$ & $-$ &$0.530\pm 0.004$&0.519 &0.523&2/3\\
\hline
$(d+\bar d)/\sum(q+\bar q)$ & $-$ &$0.368\pm 0.003$&0.370 &0.374&1/3\\
\hline
$(s+\bar s)/\sum(q+\bar q)$ & 
${{<2x\bar s(x)>}/{\sum<x(q(x)+\bar q(x))>}}$&0.090$\pm 
0.001$&0.100&0.103&0\\
&$=0.076\pm 0.022$\cite{ba95}&&&&\\ 
&$=0.10\pm 0.06$\cite{gls91}&  &  &   & \\
&$=0.15\pm 0.03$\cite{dll96}&  &  &   & \\
\hline
$(c+\bar c)/\sum(q+\bar q)$ & 0.03~\cite{bs91}$^*$&$\bf 0.011\pm 0.008$ & 
0.011&0&0\\
& 0.02~\cite{dg77}$^*$ & &  & &\\
& 0.01~\cite{gol00}$^*$ &&  & &\\
& 0.009~\cite{nnnt96}$^*$ & &  & &\\
& 0.005~\cite{hk94,pst99}$^*$& & &  & \\
& $\leq$0.004~\cite{smt99}$^*$&& & &\\
\hline
${{\sum\bar q}/{\sum q}}$ & ${{\sum<x\bar
q(x)>}/{\sum<xq(x)>}}$&$0.230\pm 0.004$&0.233&0.231&0\\
&$=0.245\pm 0.005$\cite{ba95}&&&&\\
\hline
\end{tabular}
\end{table}

\bigskip
\bigskip
\bigskip
\bigskip
\bigskip

\widetext
\begin{table}
\caption{Quark Spin Observables}
\begin{tabular}{|c|c|c|c|c|c|}\hline
\hline
Quantity & Data &SU$^{(a)}$(4)&SU$^{(b)}$(4) &SU(3)\cite{song0012}&NQM\\
\hline 
$\Delta u$ & $0.85\pm 0.04$\cite{adams97} & $0.871\pm 0.009$&0.859 
&0.863&4/3\\
$\Delta d$&$-0.41\pm$0.04\cite{adams97} &$-0.397\pm 0.002$ &$-0.393$
&$-0.397$&$-1/3$\\
$\Delta s$&$-0.07\pm$0.04\cite{adams97}&$-0.065\pm 0.000$&$-0.065$ 
&$-$0.067&0\\
\hline
$\Delta c$ & $-0.020\pm 0.004$~\cite{amt98}$^*$&$\bf -0.009\pm 0.006$ 
&$-0.009$ &0&0\\
& $-0.3$~\cite{blos98}$^*$ & & & &\\
& $-5\cdot 10^{-4}$~\cite{pst99}$^*$ & & & & \\
\hline
$\Delta\Sigma$/2 & $0.19\pm 0.06$\cite{adams97}& $0.200\pm 0.006$ 
&0.196 &0.200&1/2\\
\hline
$\Delta\bar u$, $\Delta\bar d$ & $-0.02\pm 0.11$\cite{adeva96}&0&0 &0&0\\
\hline
$\Delta\bar s$, $\Delta\bar c$ & $-$&0 &0&0&0\\
\hline
$\Delta u/\Delta\Sigma$ &$-$&$2.171\pm 0.043$&2.192 &2.162&4/3\\
$\Delta d/\Delta\Sigma$ &$-$&$-0.988\pm 0.024$&$-1.004$ &$-0.994$&$-1/3$ 
\\
$\Delta s/\Delta\Sigma$ & $-$ & $-0.162\pm 0.005$&$-0.166$ &$-0.167$&0 \\
\hline
$\Delta c/\Delta\Sigma$ & 
$-0.08\pm 0.01$~\cite{blos98}$^*$ & $\bf -0.021\pm 0.014$&$-0.022$ &0&0 
\\
  & $-0.033$ ~\cite{amt98}$^*$& & & &\\
\hline
$\Delta u/{u}$ & $-$ & $0.383\pm 0.003$&0.381 & $0.383$&2/3  \\
\hline
$\Delta d/{d}$ & $-$ & $-0.287\pm 0.001$&$-0.283$ & $-0.284$&$-1/3$  \\
\hline
$\Delta s/{s}$ & $-$ & $-3/10$ & $-0.269$ & $-0.269$&$-$  \\
\hline
$\Delta c/{c}$ & $-$ & $\bf -16/51$ &$-16/51$ & $-$&$-$  \\
\hline 
$u_\up/u_\dw$ & $-$ & $2.241\pm 0.012$&2.231 & 2.241 &5  \\
\hline
$d_\up/d_\dw$ & $-$ & $0.554\pm 0.001$&0.559 & 0.558 &1/2  \\
\hline
$s_\up/s_\dw$ & $-$ & $7/13$ & 0.576& 0.576& $-$  \\
\hline
$c_\up/c_\dw$ & $-$ & $\bf 35/67$ &35/67 & $-$ & $-$  \\
\hline
$\Gamma_1^p$ & $0.136\pm 0.016$\cite{adams97} & $0.143\pm 0.002$&0.141 
&0.142&5/18\\
$\Gamma_1^n$ & $-0.041\pm 0.007$\cite{abe97} &$-0.042\pm 0.001$&$-0.043$ 
&$-0.042$&0\\
\hline 
$a_3$&1.2670$\pm$0.0035\cite{pdg00}&$1.268\pm 0.010$&1.252 &1.260&5/3\\
$a_8$& 0.579$\pm$ 0.025\cite{pdg00}&$0.605\pm 0.006$&0.595 &0.600&1 \\
\hline
\end{tabular}
\end{table}

\widetext
\begin{table}
\caption{F, D and $G_A/G_V$ ratios of the hyperon beta decays. [a]: This 
is the best fit to four measured $(G_A/G_V)$ ($\chi^2\simeq 1.96$) under 
the constraint F+D=1.267.} 
\begin{tabular}{|c|c|c|c|c|c|}\hline
\hline
Quantity & Data &SU$^{(a)}$(4)&SU$^{(b)}$(4) &SU(3)\cite{song0012}&NQM\\
\hline 
F+D & $1.267$[a] & $1.268\pm 0.010$ &1.252& 1.260 &5/3\\
F & $0.463$[a] & $0.468\pm 0.004$ &0.462& 0.465 &2/3\\
D & $0.804$[a] & $0.800\pm 0.006$ &0.790 & 0.796 &1\\
F/D&$0.576$[a] & $0.585\pm 0.007$ &0.585 & 0.585 &2/3\\
\hline
$(G_A/G_V)_{n\to p}$&$1.2670\pm 0.0035$\cite{pdg00}&$1.268\pm 0.010$
&1.252&1.260 &5/3\\
$(G_A/G_V)_{\Lambda\to p}$&$0.718\pm 0.015$\cite{pdg00}&$0.735\pm 
0.007$&0.725& 0.730 &1\\
$(G_A/G_V)_{\Sigma^-\to n}$& $-0.340\pm 0.017$\cite{pdg00}&$-0.332\pm 
0.002$& $-0.328$ &$-0.330$ & $-1/3$\\
$(G_A/G_V)_{\Xi^-\to \Lambda}$&$0.25\pm 0.05$\cite{pdg00}&$0.202\pm 
0.002$&0.198&0.200&1/3\\
\hline
\end{tabular}
\end{table}

\begin{figure}
\begin{center}
\includegraphics[width=1.7in,angle=90]{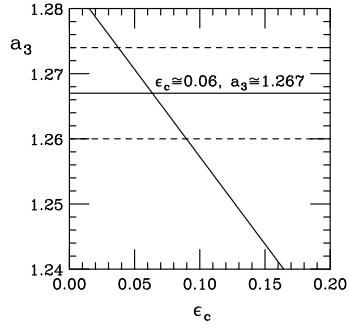}
\caption{$a_3$ as the function of $\epsilon_c$.}
\end{center}
\end{figure}

\begin{figure}
\begin{center}
\includegraphics[width=1.7in,angle=90]{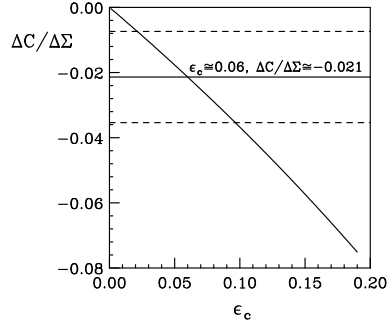}
\caption{Intrinsic charm quark polarization in the proton as the 
function of $\epsilon_c$.}
\end{center}
\end{figure}


\begin{thebibliography}{99}
\bibitem{note1} 
	Presumably one should also consider the contributions coming 
from other possible nonperturbative mechanisms, such as $gg\to q\bar q$, 
which are not included in this paper. However, it is not clear whether
these nonperturbative mechanisms are independent or not.
\end{thebibliography}
\end{document}